# A Peta-Scale Data Movement and Analysis in Data Warehouse (APSDMADW)

Ahmed Mateen
Department of Computer Science,
University of Agriculture Faisalabad, Pakistan

Lareab Chaudhary
Department of Computer Science,
University of Agriculture Faisalabad, Pakistan

## ABSTRACT
In this research paper so as to handle Information warehousing as well as on-line synthetic dispensation OLAP are necessary aspects of conclusion support, which takes more and more turn into a focal point of the data source business. This paper offers an outline of information warehousing also OLAP systems with a highlighting on their latest necessities. All of us explain backside end tackle for extract, clean-up and load information into an Data warehouse; multidimensional data model usual of OLAP; front-end user tools for query and facts evaluation server extension for useful query dispensation; and apparatus for metadata managing and for supervision the stockroom. Insights centered on complete data on customer actions manufactured goods act and souk performance are powerful advance and opposition in the internet gap .In this research, conclude the company inspiration and the program and efficiency of server's working in a data warehouse through use of some new techniques and get better and efficient results. Data in peta-byte scale. This test shows the data dropping rate in data warehouse. The locomotive is in creation at Yahoo! since 2007 and presently manages more than half a dozen peta bytes of data.

## Keywords
Data warehouse, OLAP, analytical, Map Reduce, volume, MOPS, VO, CEDPS, function, processing

## 1. INTRODUCTION
The quantity of data generate day by day in the globe is explosion. The rising size of digital and social media and internet of things be fueling it yet promote. The price of numbers growth is astounding and this data come at a velocity, with range.

Not essentially prearranged) and contain prosperity of information that preserve be a key for in advance an edging in challenging business [1].

The motivation to create MapReduce came from Google's need to process huge amount of records across a network of computers. In order to do this effectively the solution would have to handle scheduling details, while empowering the user to only write the application code for a given assignment [11].Google's MapReduce is implement in the C++ encoding verbal communication It's take a place of input proceedings with apply a map function to every of them. The map task is defined by the programmer and it outputs a list of intermediate records – the participation used for the decrease function.

### 1.1 Scheduling
- The servers that execute the processing defined by the map and decrease function be selected automatically by the middleware.
- This reduces the volume of data transfers and enables efficient processing.
- Synchronized processing: data transfers between servers for the map and reduce functions are synchronized.
- Fault tolerance: To ensure that processing can continue overall even when several servers have failed, data backups and intermediate processing results are stored automatically [2].

## 2. PREVIOUS WORK
The data movement has no authentication and reduces data. The data is not working efficiently there are much wastage of time in sending and receiving data. In servers data in a queue and prioritized data is send firstly so leas priority data some time wastage and not reached to the destination. So used new approaches to reduce the data wastage and provide efficient performance. But now the use of new techniques which is less reduces the data wastage and data surely reached at his place. Data provide High speed, Tera-scale, troubleshooting the intricate end-to-end framework, Constructing and working versatile administrations, Securing the end-to-end framework.

## 3. MATERIAL AND METHODS
### 3.1 Proposed approach
• High-speed solid information situation, to exchange information from its site of creation to different areas for ensuing examination.

• Tera-scale or speedier nearby information examination, to empower investigation of information that has been brought locally [3].

• High-execution representation, to empower scrutiny of chose subsets and elements of substantial datasets information preceding download.

• Troubleshooting the intricate end-to-end framework, which because of its bunch equipment and programming parts can come up short in an extensive variety of frequently difficult to-analyze ways[4].

Different new techniques review to get better result.

$$S_{capacity} = CSSD/dude \cdot (\lambda a + numchkpts \cdot \lambda c) \text{ titer}$$

The SSD drive itself will be picked in view of numerous components, for example, IOPS/$, GB/$, BW/$, compose perseverance, and CSSD..

$$S_{bandwidth} = N/BWPFS \cdot BW_{host2ssd}$$

In this way, considering both the limit and data transfer capacity requirements, the last arranging proportion is resolved as takes after:

$S = min\ (S_{capacity},\ S_{bandwidth})$

Given a SSD organization arrangement in view of the organizing proportion (determined as above) and a reenactment application, it might want to choose which information investigation portions can be offloaded to SSDs, without deferring the primary reproduction calculation.

$ta = \lambda a(\ 1/BWfm2c+\ 1/BWc2m+\ 1/TSSD\_k+\lambda/S/N \cdot BWPFS)$

what's more, checkpoint information yield time

$tc = \lambda c/S/N \cdot BWPFS$

In spite of the fact that the depleting of the arranged checkpoint information to the PFS can be covered with the handling of the examination information, are completed on the same SSD controller center, inferring that the checkpoint information successively [5].

$(ta + tc) \cdot S < 1$

Comprehending imbalance 6, it get the base throughput required for the examination parts that can be put on the blaze gadget:

$TSSD\_k > \lambda a \cdot S/1 − \lambda a \cdot S \cdot (\ 1/BWfm2c+ 1/BWc2m) − N \cdot (\alpha \cdot \lambda a+\lambda c)/BWPFS$

It is the whole of the vitality devoured by the SSDs amid (1) occupied time: exchanging information from register hubs to the SSD (Enode2ssd), preparing the examination information (Eactivessd), and exchanging the information to the PFS (Essd2pfs), and (2) unmoving time (Eidlessd).

$S \cdot (\lambda a+\lambda c)/BW_{host2ssd}$, where BWhost2ssd is the host to SSD interface transmission capacity. Since there are an aggregate of N S SSDs, the aggregate vitality cost for the information exchange can be communicated as takes after [6].

## 3.2. Calculation

New Data Warehouse Designing approach in view of Principal part Analysis, called DWDARPA, gets as info every one of the information set. It yields the elements outlining the most related variables, from which the information distribution center composition will be created. Truth be told, DWDARPA, an iterative procedure, works in four phases: The primary stage compresses the information for the most logical variables and studies the connection between's these variables by ascertaining the relationship lattice [7]. At the second stage, it continue the extraction of components in light of the connected variables. At this level, so compute the aggregate fluctuation mirroring the level of data that is the element including every one of the variables [8] [4].

Data – Xi…….n
Result =C :c component
Start:
// analyze Correlation betouren variables
For (J=1 to n by 1)
For (I=1 to n-1 by a) do
Corr= Calculate correlation
Store Corr(J, I)
Retrieving Factors
For (m =1 to<n by 1)
V= Commutative variance calculation
Var (m,0) = Variance Calculation
Store Var (m,o) in V
// Identification of C From V
For ( i=I to n by 1)
IF V(I,c) is max than
Put I in c
Return c
Exit

## 3.3 The Managed Object Placement Service by CEDPS

As an initial step, it's have as of late discharged a model Managed Object Placement Service (MOPS), appeared in Figure 1, which changes stockpiling into anoversaw asset. Its configuration and execution influences GridFTP, NeST, and dCache [9].

GridFTP gives an adaptable center engineering with an information interface part that permits diverse modules for included usefulness. GridFTP gives MOPS the center usefulness of quick, mass document exchanges, component 2 in this situations, which MOPS stretches out through its module capacity NeST gives ensured capacity portion by permitting the client and capacity gadget to arrange a size and term and to indicate access control records (ACLs) for record access [10] [11]. This element addresses component 3, facilitated information development, and component 4, disappointment diminishment, by diminishing the shot of plate flood mistakes [4].

dCache gives strategies to overseeing backend (tertiary) stockpiling frameworks including space administration, problem area determination, and recuperation from circle or hub disappointments. dCache likewise addresses component 3, composed information development, and component 4, disappointment decrease [12].

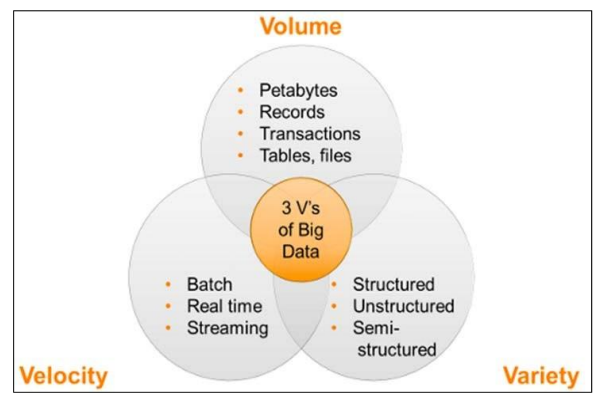

**Fig. 1: Three V's of Big Data**

## 3.4 The Data Placement Service CEDPS

CEDPS is likewise building up the Data Placement Service (DPS) that will perform information exchange operations utilizing MOPS. These objectives may be considered approaches of the work process supervisor or VO, and a strategy driven information arrangement administration is in charge of recreating and appropriating information things in conformance with these strategies or inclinations [11].

To show the adequacy of savvy information position, and coordinated the Pegasus work process administration framework from USC Information Sciences Institute with the Globus Data Replication Service, which gives proficient replication and enlistment of information sets [13] [12]. This underlying work has driven us to outline a general, offbeat Data Placement Service (DPS) that will work for the benefit of a virtual association and acknowledge information position demands from customers that reflect, for instance, gathering of documents, request of record solicitations, and so on [14]. Figure 4 outlines the operation of a DPS for stage in solicitations issued by work process administration framework information arranging demands from different contending work processes, and extra on-interest information demands from different customers.

### 3.5. Analysis

In this study we are showing how to use the datas tore and map reduce functions to process a large amount of file-based data movement and it analysis. The Map Reduce algorithm is a mainstay of many modern "big data" applications. This study operates on a single computer, but the code can scale up to use Hadoop.

Throughout this example, the data set is a collection of records for Data warehouse between 1987 and 2008. A small subset of the data set is also included with MATLAB® to allow us to run this and other examples without the entire data set.

Creating a data store allows us to access a collection of data in a chunk-based manner. A data store can process arbitrarily large amounts of data, and the data can even be spread across multiple files. We can create a data store for a collection of tabular text files, a SQL database (Database Toolbox™ required) or a Hadoop® Distributed File System (HDFS™).

ds = datastore('Filename');

dsPreview = preview(ds);

*dsPreview(:,10:15)*

The data store automatically parses the input data and makes a best guess as to the type of data in each column. In this case, use the 'Treat A sMissing' Name-Value pair argument to have datas tore replace the missing values correctly. For numeric variables (such as 'AirTime'), data store replaces every 'NA' string with a NaN value, which is the IEEE arithmetic representation for Not-a-Number.

## 4. RESULTS
### 4.1 Regression analysis of factor influenced on data wastage

The collected data regarding above factors from 5 different warehouses using secondary data approach. To check which factor is critical in data wastage at different warehouses so their applied simple liner regression. Equation is as fowling.

**Model 1:** DW= α+MSβ1+RE β2+UP β3+TS β4+SS β5+DTβ6+ε

**Model 2:** DW= α+OISβ1+PISβ2+TISβ3+ε

**Model 3:** DW= α+DQβ1+SQ β2+ε

Encoded organization with 1 if they have above feature and 0 for those don't have above factors.

In Model 1 Data wastage (DW) is dependent variable and Management support (MS), Resources (RE), User Perception (UP), Team Skills (TS), Source System (SS) and Development Technology (DT) are independent variables.

Development technology is also positive relation with data wastage as line touches more points [15]. Therefore data wastage in warehouse is more dependent on technology used while peta-scale data is design and developed. As per the influence of implementation factors the Multiple R represented that it is up to 50%. Means that implementation factors have 50% influence on data wastage in Data warehouses [16].

**Table 1. Data of various Servers**

| Server Num | Tail Num | Actual Elapsed Time | CRS Elapsed Time | Extra Time | Delay |
|---|---|---|---|---|---|
| 1503 | 'NA' | 53 | 57 | 'NA' | 8 |
| 1550 | 'NA' | 63 | 56 | 'NA' | 8 |
| 1589 | 'NA' | 83 | 82 | 'NA' | 21 |
| 1655 | 'NA' | 59 | 58 | 'NA' | 13 |
| 1702 | 'NA' | 77 | 72 | 'NA' | 4 |
| 1729 | 'NA' | 61 | 65 | 'NA' | 59 |
| 1763 | 'NA' | 84 | 79 | 'NA' | 3 |
| 1800 | 'NA' | 155 | 143 | 'NA' | 11 |

**Table 2. Model 1**

| Regression Statistics | |
|---|---|
| Multiple R | 0.492366 |
| R Square | 0.242424 |
| Adjusted R Square | -0.76364 |
| Standard Error | 3.535534 |
| Observations | 10 |

**Table 3. ANOVA Table**

| ANOVA | Df | SS | MS | F | Significance F |
|---|---|---|---|---|---|
| Regression | 6 | 20 | 3.333333333 | 0.4 | 0.8435099 |
| Residual | 5 | 62.5 | 12.5 | | |
| Total | 11 | 82.5 | | | |

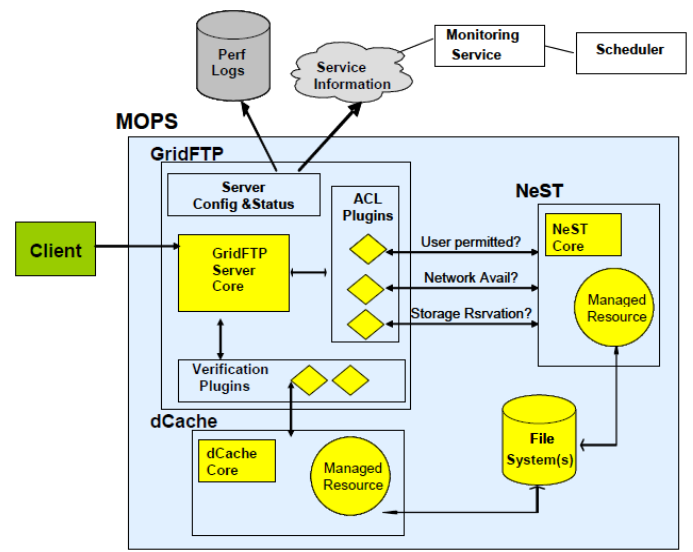

**Fig. 3: General MOPS Architecture**

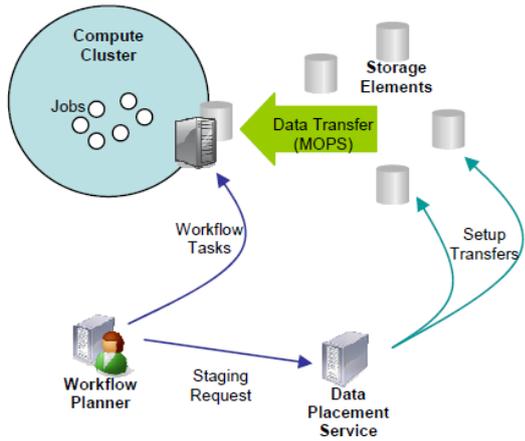

**Fig. 4: Workflow Management**

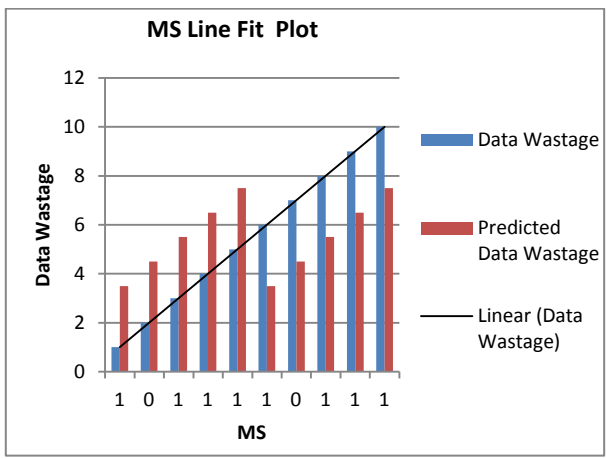

**Fig. 5: Data Wastage and Management Support**

Above figure shows that data wastage is particularly dependent upon management support as line touching almost every data set.

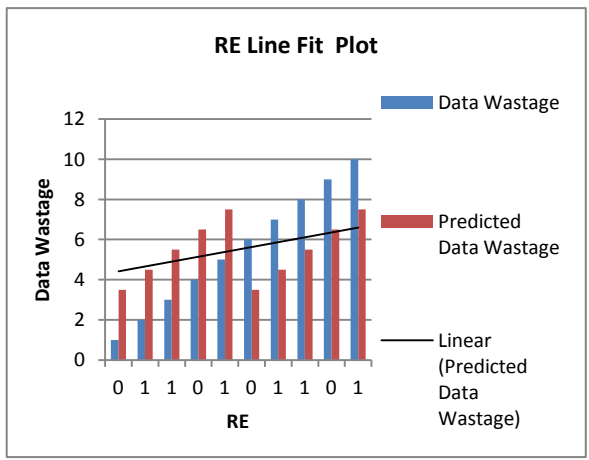

**Fig. 6: Data Wastage and Resources**

Above figure shows that data wastage is partially dependent upon Resources as line not said to be good fit.

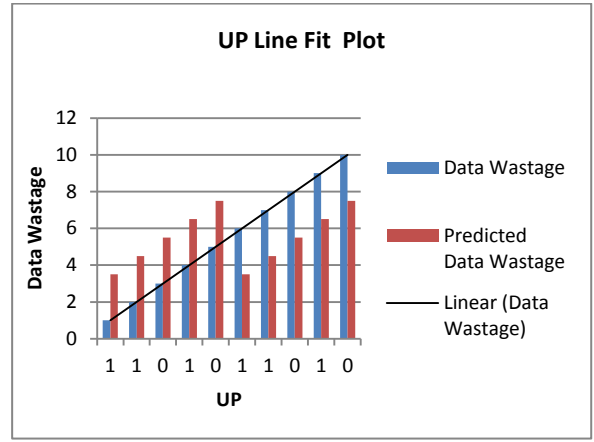

**Fig. 7: Data Wastage and User Participation**

Above figure shows that data wastage is more dependent upon user perception as it is touching most of points and said to be good fit.

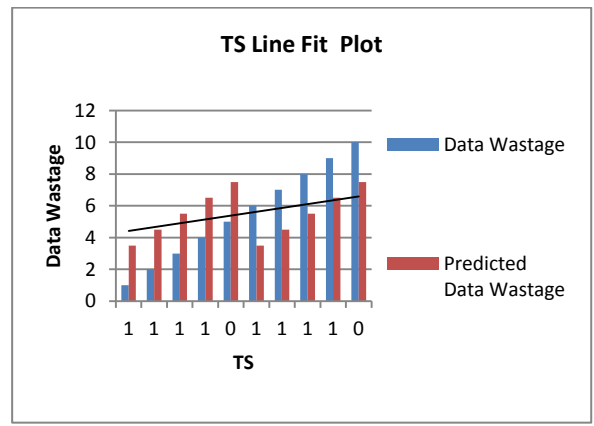

**Fig. 8: Data Wastage and Team Skills**

Above figure shows that data wastage is said to be partially dependent upon skills as line less good fits.

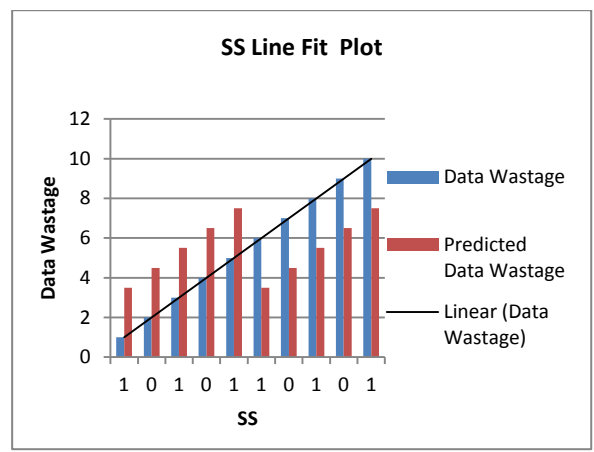

**Fig. 9: Data Wastage and Source System**

Above figure source system have more influence on data wastage as figure shows more points are touching the line.

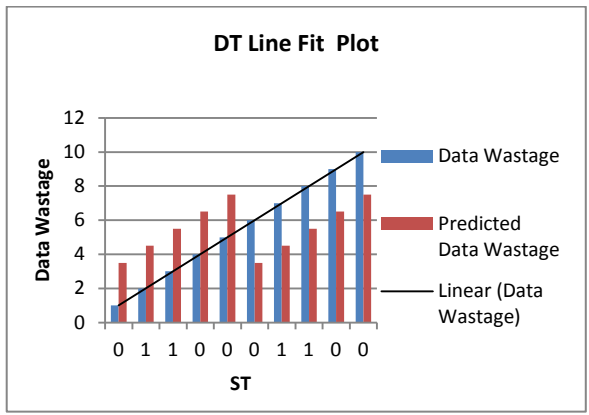

**Fig. 10: Data Wastage and Development Technology**

## 5. CONCLUSION AND FURTURE WORK

This paper presented the SciDAC Center for Enabling Distributed Peta-scale Science (CEDPS), which is tending to three issues basic to empowering the conveyed administration and examination of peta-scale datasets: information arrangement, adaptable administrations, and investigating. In information arrangement, it is created apparatuses and procedures for dependable, superior, secure, and strategy driven situation of information inside an appropriated science environment. There is built an oversaw object situation administration (MOPS)— a noteworthy upgrade to today's Grid FTP—that takes into account administration of the space, transmission capacity, associations, and different assets expected to exchange information to and/or from a capacity framework.

It developed a conclusion to-end checking design that utilizations instrumented administrations to give point by point information to both foundation accumulation and run-time occasion driven gathering.

In future proposed technique will be implemented in different projects according to the specified approach so better statistics and results will be suggest in term of big data environments so data efficiency and time will be managed more efficiently .